\documentclass[intlimits,twoside,a4paper]{article}

\usepackage{amsmath,amssymb}
\usepackage{graphicx}
\usepackage{wrapfig}

\usepackage[T2A]{fontenc}
\usepackage[cp1251]{inputenc}
\usepackage[eqsecnum]{cmpj2}

\issue{2012}{15}{3}{33602}

\doinumber{10.5488/CMP.15.33602}

\title[Effect of distribution of stickers along backbone]%
{Effect of distribution of stickers along backbone on
temperature-dependent
structural properties in associative polymer solutions%
}
\author[X.-G. Han \textsl{et al.}]{X.-G. Han\footnote{E-mail: xghan0@163.com
} , X.-F. Zhang, Y.-H. Ma}
\address{
School of Mathematics, Physics and Biology, Inmongolia
Science and Technology University, \\ Baotou 014010, China }

\authorcopyright{X.-G. Han, X.-F. Zhang, Y.-H. Ma, 2012}

\date{Received March 29, 2012,  in final form June 8, 2012}

\begin{document}

\maketitle

\begin{abstract}
Effect of distribution of stickers along the backbone on structural
properties in associating polymer solutions is studied using
self-consistent field lattice model. Only two inhomogeneous
morphologies, i.e., microfluctuation homogenous (MFH) and micelle
morphologies, are observed. If the system is cooled, the solvent
content within the aggregates decreases. When the spacing of stickers
along the backbone is increased the temperature-dependent range of
aggregation in MFH morphology and half-width of specific heat peak
for homogenous solutions-MFH transition increase, and the symmetry
of the peak decreases. However, with increasing spacing of stickers,
the above three corresponding quantities related to micelles behave
differently. It is demonstrated that the broad nature of the
observed transitions can be ascribed to the structural changes which
accompany the replacement of solvents in aggregates by polymer,
which is consistent with the experimental conclusion. It is found
that different effect of spacing of stickers on the two transitions
can be interpreted in terms of intrachain and interchain
associations.

\end{abstract}

\section{Introduction}

Physically associating polymers are polymer chains containing a
small fraction of attractive groups along the backbones. The
attractive groups tend to form physical links which can play a
important role in reversible junctions between different polymer
chains. The junctions can be broken and recombined frequently on
experimental time scales. This property of the junctions makes
associative polymer solutions behave reversibly when ambient
conditions, such as temperature and concentrations, change. This
tunable characteristic of the system produces extensive applications~\cite{Clar1987,Slat1998,Tong2001,Tayl1998} that have a great potential
as smart materials~\cite{Brun2001,Gree2008,Cord2008}.

Physically associating polymer is simply considered as an
amphiphilic block copolymer. When dissolved in a solvent such as
water, amphiphilic copolymers can self-assemble into micelles. The
solvophobic blocks (attractive groups) cluster together to form a
core, and the solvophilic blocks spread outward as a corona. One key
effect of solvent selectivity (or equivalently, temperature) is that
the micelles dissolve into single chains at a critical micelle
temperature as the solvent selectivity decreases. Several studies
have investigated the detailed structure of micelles with varying
temperature in aqueous
solutions~\cite{Pede2003,Somm2004,Cast2004,Bang2004} One appealing
advantage of polymers is their versatility. Architectural parameters
of associative polymers may be tuned by changing the chain length,
chemical composition and distribution of attractive
groups~\cite{Chas1998,Beau2002,Beau2002a,Lafl2003,Lafl2003a,Brow1992,
Gind2008}. It was suggested that the distribution of stickers along
the chains can be an important factor in controlling macroscopic
properties of these systems~\cite{Brow1992, Gind2008}. The study of
the effect of distribution of sticker along the chain in physically
association polymer solutions (PAPSs) would be useful to establish
and understand the thermodynamics of block copolymers in a selective
solvent.

 It is well-known that self-consistent field theory (SCFT), as a
mean-field theory, has been applied to the study of a great deal of
problems in polymeric
systems~\cite{Orland1996,Mats1994,Tang2004,He2004,Chen2006}.
Recently, SCFT is applied to the study of the properties of micelles in
polymer solutions~\cite{Cava2006,Jeli2007,Char2008}. In previous
paper~\cite{Han2010}, we focused on the thermodynamic properties and
structure transitions in PAPSs. The microfluctuation homogenous
(MFH) and  micelle morphologies were observed. The degrees of
aggregation of micelle morphology is much larger than that of MFH
morphology. In this work, the effect of the distribution of stickers
along the backbone on structures in PAPSs is studied using
a self-consistent field lattice model. The temperatures at which the
above two inhomogenous morphologies first appear, denoted by
$T_{\mathrm{MFH}}$ and $T_\mathrm{m}$, respectively, are dependent on the
spacing of sticker along the chain. If the system is cooled from
$T_{\mathrm{MFH}}$ and $T_\mathrm{m}$, the solvent content within the aggregates
(microfluctuation or micellar core) decreases, which is dependent on
spacing of sticker and morphology. The increase in spacing of
sticker has different effect on homogenous solutions-MFH and
MFH-micelle transitions. It is
 found that this result can be interpreted in
terms of intrachain and interchain associations.

\section{Theory\label{sec2}}

We consider a system of incompressible PAPSs, where $n_\mathrm{P}$
polymers are each composed of $N_\mathrm{st}$ segments of type sticker monomer
(attractive group) and $N_\mathrm{ns}$ segments of type nonsticky monomer,
distributed over a lattice. The sticker monomers are placed at the two ends of a chain and regularly along the chain backbone, and there are $l$ nonsticky monomers between two neighboring sticker monomers. The degree of
polymerization of chain is $N=N_\mathrm{st}+N_\mathrm{ns}$. In addition to
polymer monomers, $n_\mathrm{h}$ solvent molecules are placed on the vacant lattice sites.
Stickers, nonsticky monomers and solvent molecules have the same
size and each occupies one
lattice site. The total number of lattice sites is $N_{L}=n_\mathrm{h}+n_\mathrm{P}N$. Nearest neighbor pairs of stickers have attractive
interaction $-\epsilon$ with $\epsilon>0$, which is the only
non-bonded interaction in the present system. In this simulation,
however, instead of directly using the exact expression of the
nearest neighbor interaction for stickers, we introduce a local
concentration approximation for the non-bonded interaction similar
to the references~\cite{Han2010,Chen2006}.
The interaction energy is expressed as:%
\begin{equation}
\frac{U}{k_\mathrm{B}T}=-\chi \sum_{{r}}\widehat{\phi }_\mathrm{st}({r})\widehat{\phi }%
_\mathrm{st}({r}),  \label{01}
\end{equation}%
where $\chi $ is the Flory-Huggins interaction parameter in the
solutions, which equals $\frac{z}{2k_\mathrm{B}T}\epsilon$, $z$ is the
coordination number of the lattice used, where $\sum_{r}$ means the
summation over all the lattice sites ${r}$ and $\widehat{\phi
}_\mathrm{st}({r})=\sum_{{j}}\sum_{s{\in \mathrm{st}}}\delta
_{{r},{r}_{j,s}}$ is the volume fraction of stickers on site ${r}$, where $j$ and $%
s$ are the indexes of chain and monomer of a polymer, respectively.
$s\in \mathrm{st}$ means that the $s$th monomer belongs to sticker
monomer type. We perform the SCFT calculations in the canonical ensemble,
and the field-theoretic free energy $F$ is defined as

\begin{equation}
\frac{F[\omega _{+},\omega _{-}]}{k_\mathrm{B}T}=\sum_{r}\left[ \frac{1}{4\chi }%
\omega _{-}^{2}(r)-\omega _{+}(r)\right] -n_\mathrm{P}\ln Q_\mathrm{P}[\omega
_\mathrm{st},\omega _\mathrm{ns}]-n_\mathrm{h}\ln Q_\mathrm{h}[\omega _\mathrm{h}],  \label{free0}
\end{equation}%
where $Q_{\mathrm{h}}$ is the partition function of a solvent molecule
subject to the field $\omega _\mathrm{h}(r)=\omega _{{+}}(r)$, which
is defined as $Q_\mathrm{h}=\frac{1}{n_\mathrm{h}}\sum_{r}\exp\left[ - \omega
_\mathrm{h}(r)\right]$. $Q_\mathrm{{P}}$ is the partition function of a
noninteracting polymer chain subject to the fields $\omega
_\mathrm{st}(r)=\omega _{{+}}(r)-\omega _{{-}}(r)$ and $\omega
_\mathrm{ns}(r)=\omega _{{+}}(r)$, which act on sticker and nonsticky
segments, respectively.

Equation~(\ref{free0}) can be considered an
alternative form of the self-consistent field
free energy functional for an incompressible polymer solutions~\cite{Fredr2005}.
When a local concentration approximation for the
non-bonded interaction is introduced, the SCFT descriptions of
lattice model for PAPSs presented in this work is basically
equivalent to that of the ``Gaussian thread model'' chain for the
similar polymer solutions~\cite{Fredr2005}. The related illumination
in detail refers to reference~\cite{Han2010}.

 Minimizing the free energy function $F$ with
$\omega _{-}(r)$ and $\omega _{+}(r)$ leads to the following
saddle point equations:
\begin{equation}
\omega _{-}(r)=2\chi \phi _\mathrm{st}(r),  \label{scf1}
\end{equation}%
\begin{equation}
\phi _\mathrm{st}(r)+\phi _\mathrm{ns}(r)+\phi _{h}(r)=1, \end{equation} where
\begin{equation}
\phi _\mathrm{st}(r)=\frac{1}{N_{L}}\frac{1}{z}\frac{n_\mathrm{P}}{Q_\mathrm{P}}\sum_{s\in \mathrm{st}%
}\sum_{\alpha _{s}}\frac{G^{\alpha _{s}}(r,s|1)G^{\alpha _{s}}(r,s|N)}{G(r,s)%
}
\end{equation}
and
\begin{equation}
\phi _\mathrm{ns}(r)=\frac{1}{N_{L}}\frac{1}{z}\frac{n_\mathrm{P}}{Q_\mathrm{P}}\sum_{s\in \mathrm{ns}%
}\sum_{\alpha _{s}}\frac{G^{\alpha _{s}}(r,s|1)G^{\alpha _{s}}(r,s|N)}{G(r,s)%
}
\end{equation}
are the average numbers of sticker and nonsticky segments at $r$,
respectively, and
\[
\phi_\mathrm{h}(r)=\frac{1}{N_{_{L}}}\frac{n_\mathrm{h}}{\
Q_{_\mathrm{h}}}\exp \left\{ -\ \omega _\mathrm{h}(r)\right\}
\]
 is the average
numbers of solvent molecules at $r$. $Q_\mathrm{P}$ is expressed as
\[
Q_\mathrm{P}=\frac{1%
}{N_{L}}\frac{1}{z}\sum_{r_{N}}\sum_{{\alpha }_{N}}G^{\alpha _{N}}(r,N|1),
\]
where $r_{N}$ and ${\alpha }_{N}$ denote the position and
orientation of the $N$th segment of the chain, respectively.
$\sum_{r_{N}}\sum_{\alpha _{N}}$ means the summation over all the
possible positions and orientations of the $N$th segment of the
chain. $G^{\alpha _{s}}(r,s|1)$ and $G^{\alpha _{s}}(r,s|N)$ are the
end segment distribution functions of the $s$th segment of the
chain. $G(r,s)$ is the free segment weighting factor. The
expressions of the above three quantities refer to Appendix. In this
work, the chain is described as a random walk without the
possibility of direct backfolding. Although self-intersections of a
chain are not permitted, the excluded volume effect is sufficiently
taken into account~\cite{Medv2001}.

The saddle point is calculated using the pseudo-dynamical evolution
process~\cite{Han2010}.
The calculation is initiated from appropriately random-chosen fields $%
\omega_{+}(r)$ and $\omega _{-}(r)$, and interrupted when the change
of free energy $F$ between two successive iterations is reduced to
the needed precision. The resulting configuration is taken as a
saddle point one. By comparing the free energies of the saddle point
configurations obtained from different initial fields, the relative
stability of the observed morphologies can be assessed.

\section{Result and discussion\label{sec3}}

In our studies, the properties of associative polymer solutions
depend on four tunable parameters: $\chi$ is the Flory-Huggins
interaction parameter, $N$ is the degree of polymerization of chain,
where $N$ equals 81 in this paper, $l$ is the spacing of stickers along the
backbone and $\bar{\phi}_\mathrm{st}$ is the average volume fraction of
polymers. The calculations are performed in a three-dimensional simple
cubic lattice with periodic boundary condition, and the effect of
the lattice size is considered. The results presented below are
obtained from the lattice with $N_{L}=40^{3}$. Three different
morphologies in PAPSs are observed, i.e., the homogenous,
micro-fluctuation homogenous (MFH) and micelle  morphologies. By
comparing the relative stability of the observed states, the phase
diagram is constructed.

\begin{figure}[ht]
\centerline{
\includegraphics[width=0.48\textwidth]{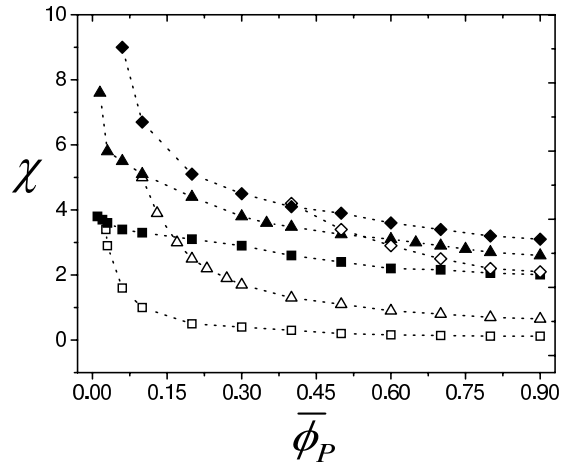}
}
\caption{The
phase diagram for systems with different spacing of stickers
$l$. The boundaries of MFH and micelle morphologies are obtained.
The red open and solid squares, green open and solid triangles, blue
open and solid diamonds correspond to the boundaries of MFH and
micelle morphologies for $l=3, 9, 19$, respectively.\label{phadia}}
\end{figure}
Figure~\ref{phadia} shows the phase diagram of the systems with
different spacing of stickers $l$. At fixed $l$, when $\chi$ is
increased from homogenous solutions, MFH and micelle morphologies
appear in turn. $\phi _\mathrm{P}(r)$ and $\phi _\mathrm{st}(r)$ in the
solutions with MFH morphology slightly fluctuate around
$\bar{\phi}_\mathrm{P}$ and $\bar{\phi}_\mathrm{st}$, respectively. The average
volume fraction of stickers at the sticker-rich sites (fluctuations)
$r_\mathrm{ri}$ increases with increasing $\chi $ for fixed
$\bar{\phi}_\mathrm{P}$. Its maximum value is about $2\bar{\phi}_\mathrm{st}$,
which is much smaller than unity for all the $\bar{\phi}_\mathrm{P}$.
There exists the state of microfluctuations whose thermodynamics is adequately
captured by SCFT. It is confirmed that the MFH appearance is accompanied
by the appearance of the heat capacity peak (shown below), which is in
reasonable agreement with the conclusion drawn by Kumar et
al.~\cite{Han2010,Kumar2001}. The basic component of micelle
morphology is flower micelles, which are randomly and closely distributed
in the system. Each micelle has a sticker-rich core, which is
located at the center of a micelle, surrounded by non-sticky
components of polymers. The average value of volume fraction of
stickers at the micellar core is much larger than that of
sticker-rich sites in MFH morphology. It is shown that the degree of
aggregation of stickers in micelle morphology is much larger than
that in MFH morphology.

When spacing of stickers $l$ is changed, only MFH morphology and
micelles are observed as inhomogeneous morphologies. The structural
morphology of MFH morphology does not change, and the micellar shape
remains spherelike. For $l=19$, the $\chi$ value on micellar
boundary ($\sim{1}/{T_\mathrm{m}}$) increases with decreasing
$\bar{\phi}_\mathrm{P}$. When $\bar{\phi}_\mathrm{P}$ goes down to a certain
extent, micellar boundary becomes steep. The  $\chi$ value on MFH
boundary ($\sim{1}/{T_\mathrm{MFH}}$) also rises with a decrease in
$\bar{\phi}_\mathrm{P}$. MFH boundary intersects the micellar one at
$\bar{\phi}_\mathrm{CFC}$, which is the critical MFH concentration
($\bar{\phi}_\mathrm{CFC}=0.4$). When $l$ is decreased, at fixed
$\bar{\phi}_\mathrm{P}$, the $\chi$ value on micellar boundary shifts to
a small value, and the $\chi$ value on MFH boundary decreases
markedly. $\bar{\phi}_\mathrm{CFC}$ also drops with a decrease in $l$.

In this paper, the structural properties dependent on temperature
are focused. Therefore, the quantities related to volume fractions
of stickers in MFH and micelle morphologies as a function of $\chi$
are studied.
\begin{figure}[h]
\includegraphics[width=0.48\textwidth]{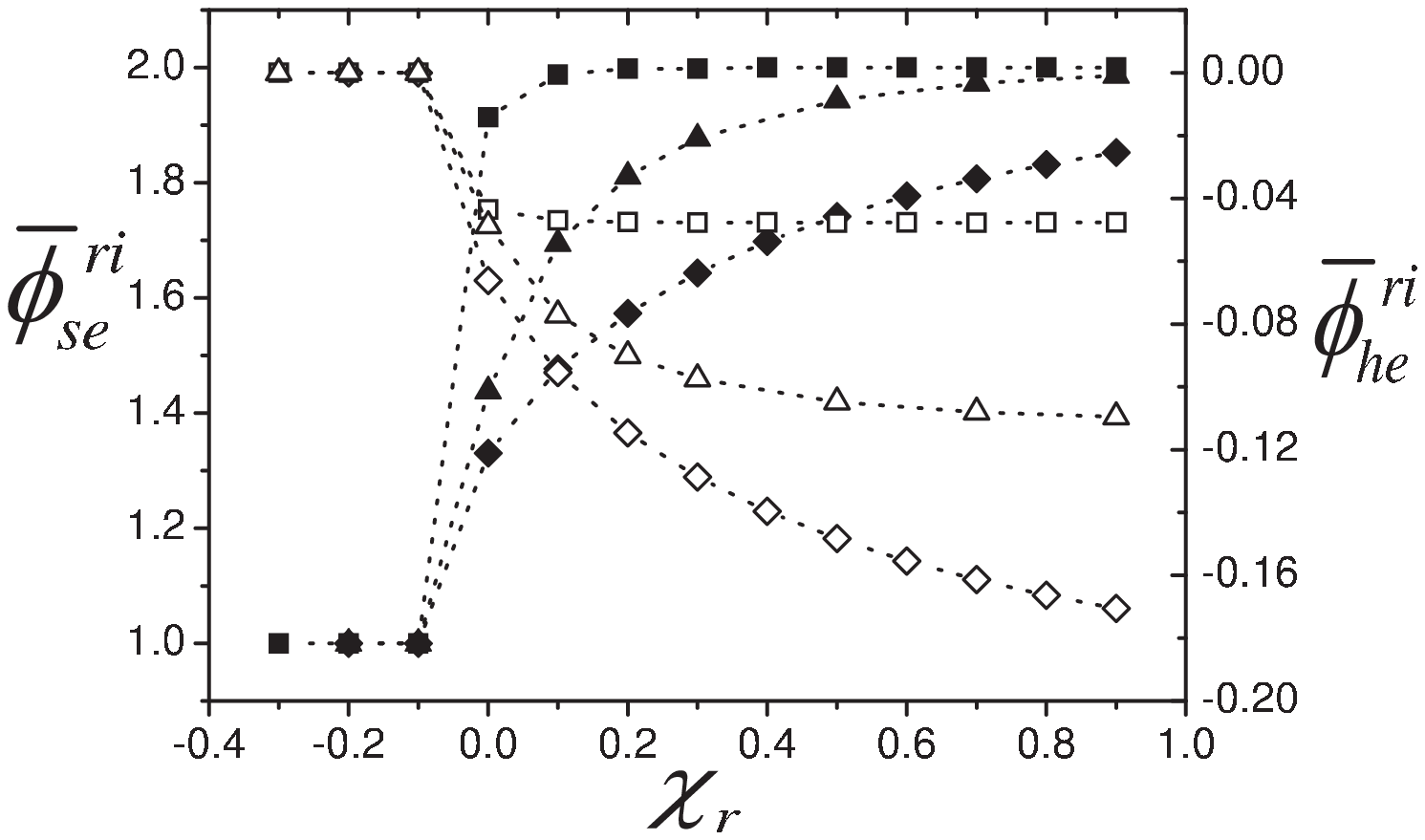}%
\hfill%
\includegraphics[width=0.48\textwidth]{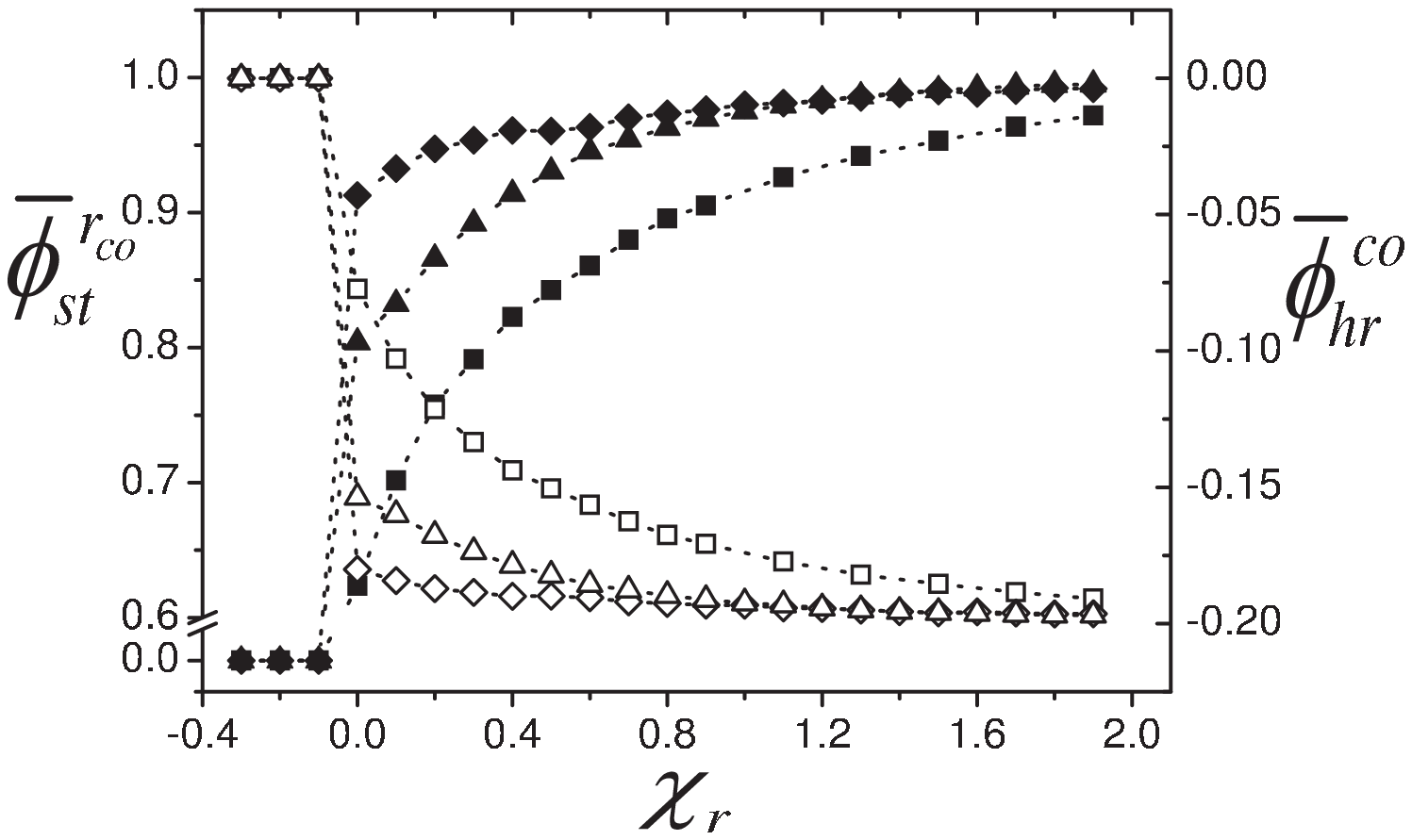}%
\\%
\parbox[t]{0.48\textwidth}{%
\centerline{(a)}%
}%
\hfill%
\parbox[t]{0.48\textwidth}{%
\centerline{(b)}%
}%
\caption{The variations of effective average volume fractions of
stickers  and solvents at the sticker-rich sites in MFH morphologies
with different spacing of stickers $l$, denoted by $\bar{\phi}
_\mathrm{se}^\mathrm{ri}\left[=\bar{\phi}_\mathrm{st}^\mathrm{ri}\big/\bar{\phi}_\mathrm{st}\right]$  and
$\bar{\phi}^\mathrm{ri}_\mathrm{he}\left[=\left(\bar{\phi}^\mathrm{ri}_\mathrm{h}-\bar{\phi}_\mathrm{h}\right)
\big/\bar{\phi}_\mathrm{st}\right]$, respectively, with the $\chi$ deviation from MFH boundary $\chi_\mathrm{r}$
at $\bar{\phi}_\mathrm{P}=0.8$ are presented in figure~\ref{aggr}~(a); The
variations of the average volume fraction of stickers  and relative
average volume fraction of solvents at the micellar cores, denoted
by $\bar{\phi} _\mathrm{st}^{r_\mathrm{co}}$ and $\bar{\phi}^{r_\mathrm{co}}_\mathrm{hr}
\left[=\bar{\phi}^{r_\mathrm{co}}_\mathrm{h}-\bar{\phi}_\mathrm{h}\right]$, respectively, with the
$\chi$ deviation from micellar boundary $\chi_\mathrm{r}$ in the systems with
different spacing of stickers $l$ at $\bar{\phi}_\mathrm{P}=0.8$ are
shown in figure~\ref{aggr}~(b).} \label{aggr}
\end{figure}
Figure~\ref{aggr}~(a) shows the variations of effective
average volume fractions of stickers and solvents at the
sticker-rich sites (microfluctuations) in MFH morphologies with
different spacing of stickers $l$, which are denoted by $\bar{\phi}
_\mathrm{se}^\mathrm{ri}$ and $\bar{\phi}^\mathrm{ri}_\mathrm{he}$, respectively, with the
$\chi$ deviation from MFH boundary, $\chi_\mathrm{r}$, at
$\bar{\phi}_\mathrm{P}=0.8$, where $\bar{\phi} _\mathrm{se}^\mathrm{ri}$ and
$\bar{\phi}^\mathrm{ri}_\mathrm{he}$ equal $\bar{\phi}_\mathrm{st}^\mathrm{ri}/\bar{\phi}_\mathrm{st}$
and $(\bar{\phi}^\mathrm{ri}_\mathrm{h}-\bar{\phi}_\mathrm{h})/\bar{\phi}_\mathrm{st}$,
respectively. With the increase in $\chi_\mathrm{r}$, $\bar{\phi} _\mathrm{se}^\mathrm{ri}$
at $l=3$ initially rises when $0\leqslant\chi_\mathrm{r}\leqslant0.2$ and then
maintains a certain value, and the corresponding
$\bar{\phi}_\mathrm{he}^\mathrm{ri}$ first decreases  when $0\leqslant\chi_\mathrm{r}\leqslant0.2$
and then remains constant. Although the increase of the degree of
aggregation in MFH morphology is accompanied by the penetration of
solvents, the effective total quantity of penetration of solvents is
very small ($|\bar{\phi}_\mathrm{he}^\mathrm{ri}|=0.047$). When $l$ is increased,
the shapes of the curves of $\bar{\phi} _\mathrm{se}^\mathrm{ri}(\chi_\mathrm{r})$ and
$\bar{\phi}_\mathrm{he}^\mathrm{ri}(\chi_\mathrm{r})$ are similar to the case of $l=3$.
However, the onset of the range independent of $\chi_\mathrm{r}$ shifts to
a larger $\chi_\mathrm{r}$ value. The minimum of $\bar{\phi}_\mathrm{he}^\mathrm{ri}(\chi_\mathrm{r})$
goes down with an increasing $l$. It is demonstrated that in MFH
morphology the increase in spacing of stickers augments the
temperature-dependent range of aggregation of stickers and
accelerates the effective penetration of solvents.

The variations of the average volume fraction of stickers  and the relative
average volume fraction of solvents at the micellar cores, which are
denoted by $\bar{\phi} _\mathrm{st}^{r_\mathrm{co}}$ and
$\bar{\phi}^{r_\mathrm{co}}_\mathrm{hr}$
(=$\bar{\phi}^{r_\mathrm{co}}_\mathrm{h}-\bar{\phi}_\mathrm{h}$), respectively, with the
$\chi$ deviation from micellar boundary, $\chi_\mathrm{r}$, in the systems
with different spacings of stickers $l$ at $\bar{\phi}_\mathrm{P}=0.8$
are shown in figure~\ref{aggr}~(b). At $l=3$, $\bar{\phi}
_\mathrm{st}^{r_\mathrm{co}}$ rises and approaches 1, and the corresponding
$\bar{\phi}^{r_\mathrm{co}}_\mathrm{hr}$ decreases and is close to its minimum
when $\chi_\mathrm{r}$ is increased. When $l$ is increased, $\bar{\phi}
_\mathrm{st}^{r_\mathrm{co}}$ goes up and rapidly approaches 1, and the
corresponding $\bar{\phi}^{r_\mathrm{co}}_\mathrm{hr}$ goes down and is quickly
close to its minimum, with increasing $\chi_\mathrm{r}$. It is demonstrated
that in micelle morphology, the increase in spacing of stickers almost
does not change the total quantity of the expelled solvent, and
decreases the effective range of aggregation dependent on $\chi_\mathrm{r}$,
which is contrary to that of MFH morphology.

\begin{figure}[h]
\includegraphics[width=0.48\textwidth]{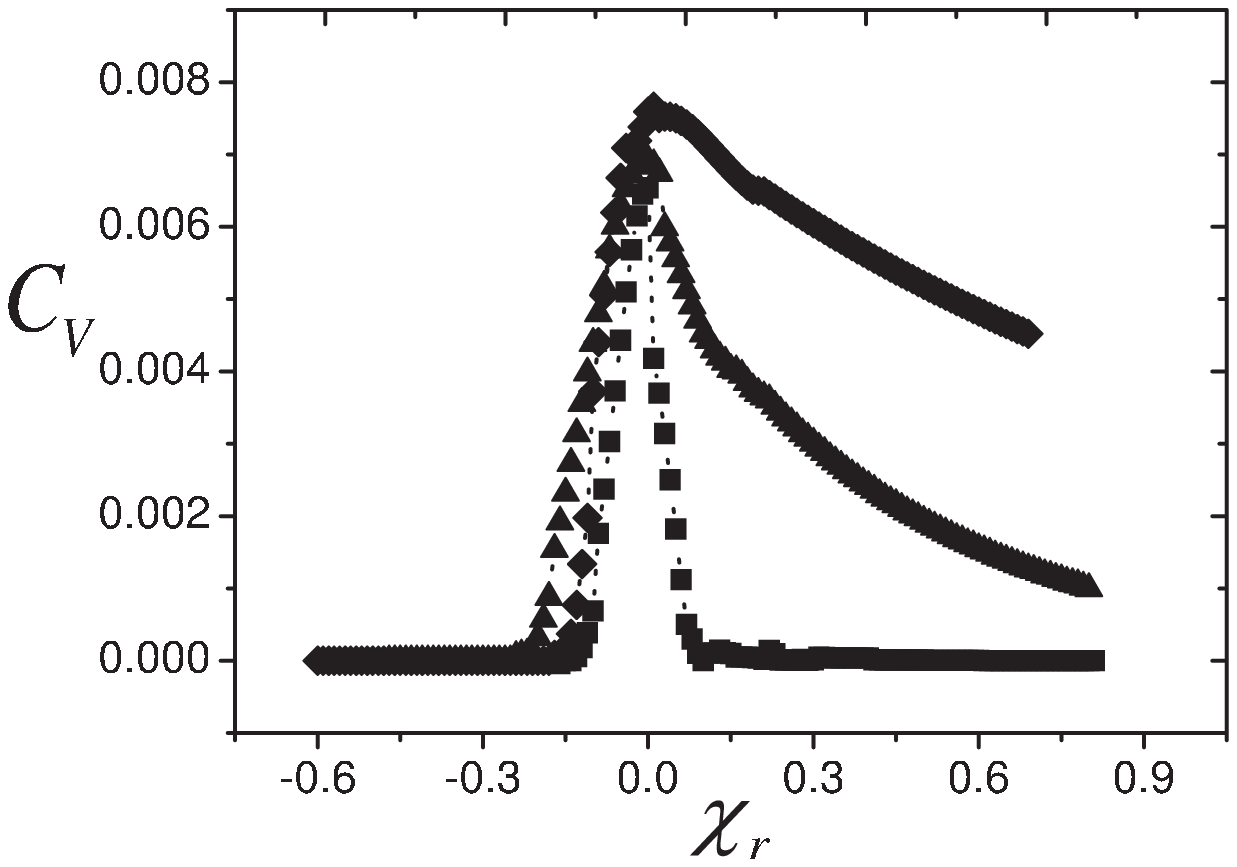}%
\hfill%
\includegraphics[width=0.48\textwidth]{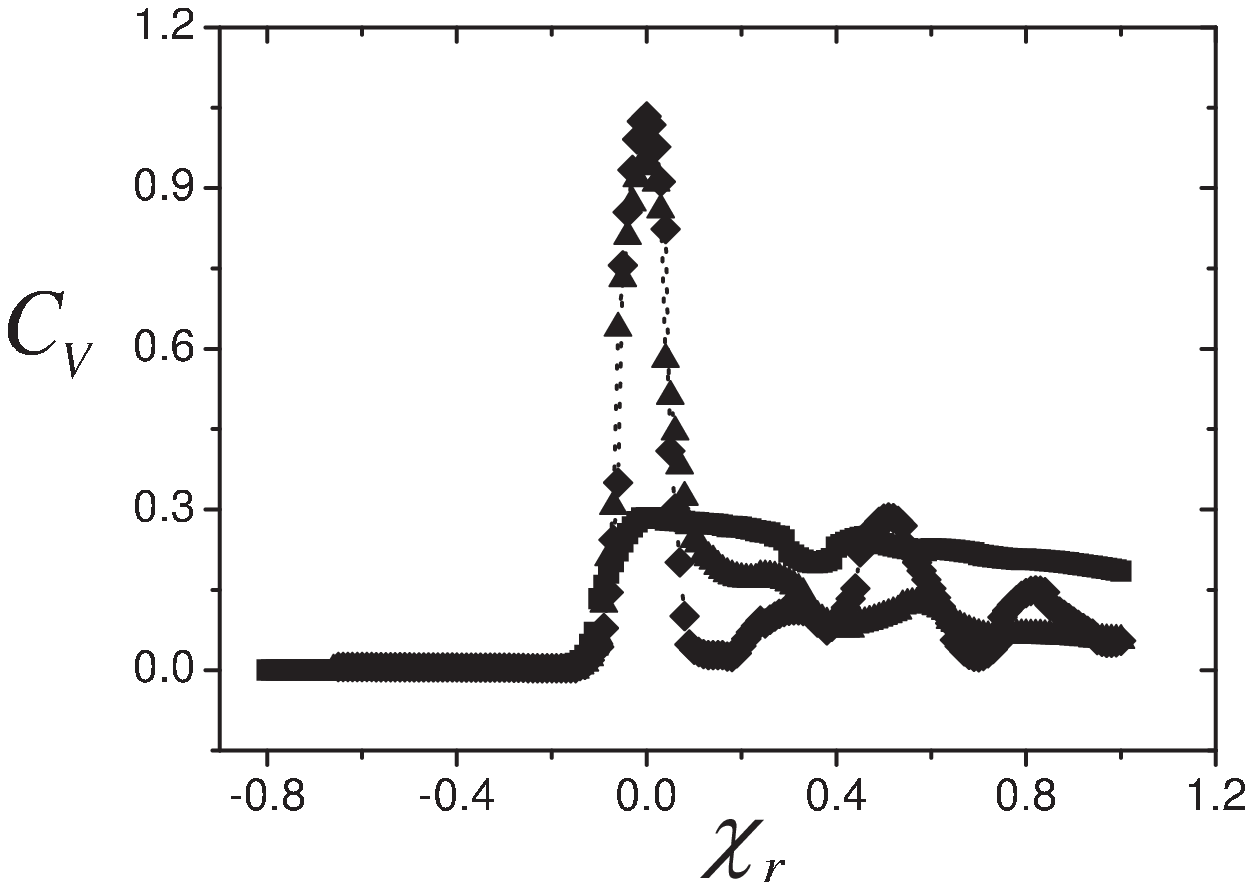}%
\\%
\parbox[t]{0.48\textwidth}{%
\centerline{(a)}%
}%
\hfill%
\parbox[t]{0.48\textwidth}{%
\centerline{(b)}%
}%
\caption{(Color online) The variations of the specific heat $C_{\mathrm{V}}$ with the
$\chi$ deviation from MFH or micellar boundary $\chi_\mathrm{r}$ in HS-MFH
and MFH-micelle transition regions at $\bar{\phi}_\mathrm{P}=0.8$ in the
systems with different spacing of stickers $l$ are presented in
figures~\ref{Cv}~(a) and (b), respectively. The red squares, green
triangles and blue diamonds correspond to $l=3, 9, 19$,
respectively.} \label{Cv}
\end{figure}
In order to demonstrate the property of the observed transition, the
heat capacity $C_{\mathrm{V}}$ is calculated, because the half-width of a specific
heat peak may be an intrinsic measure of transition
broadness~\cite{Doug2006}. In this work, the heat capacity per site
of PAPSs is expressed as follows (in the unit of $k{_\mathrm{B}}$):
\begin{eqnarray} C_{\mathrm{V}} &=&\left(\frac{\partial {U}}{\partial
{T}}\right)_{N_{L},n_\mathrm{P}}
\label{scf5-2} \nonumber\\
&=&\frac{1}{N_{L}}\chi ^{2}\frac{\partial }{\partial {\chi
}}\left( \sum_{r}\phi _\mathrm{st}^{2}(r)\right) .
\end{eqnarray}
The $C_{\mathrm{V}}(\chi_\mathrm{r})$ curves for the HS-MFH and MFH-micelle
transitions in various $l$ at $\bar{\phi}_\mathrm{P}=0.8$ are shown in
figures~\ref{Cv}~(a) and \ref{Cv}~(b), respectively. For HS-MFH
transition, a peak appears in each $C_{\mathrm{V}}(\chi_\mathrm{r}  )$ curve. When
$l$ is increased, the height and half-width of the transition peak
rise, and the symmetry of the peak decreases. Meanwhile, for the
MFH-micelle transition, there are some peaks in each  $C_{\mathrm{V}}(\chi_\mathrm{r})$
curve.  The highest of these peaks, corresponds to
MFH-micelle transition. When $l$ is increased, the height of the
transition peak rises, and the corresponding half-width does go
down. The shape of the transition peak tends to be symmetric under
the condition of an increasing $l$. It is shown that an increase in
$l$ causes an increase in the broadness of HS-MFH transition
and a decrease in the broadness of MFH-micelle transition. The effect of
spacing of stickers on $C_{\mathrm{V}}(\chi_\mathrm{r} )$ curve for the MFH-micelle
transition is different from that for HS-MFH transition.

\looseness=-1The HS-MFH transition that took place in this work, which
corresponds to the clustering transition observed by Kumar et
al.~\cite{Kumar2001,Han2010}, is affected by the change of spacing of
stickers as discussed above. When spacing of stickers is increased,
both the magnitude of the temperature-dependent range of aggregation
and the effective total quantity of the expelled solvents in MFH
morphology increase. Under the same condition, the broadness of
the corresponding transition also increases. Meanwhile, for MFH-micelle
transition, an increase of spacing of stickers results in a
decrease of the magnitude of the effective range dependent on
temperature and on the transition broadness. Overall, for the above two
transitions, the magnitude of the temperature-dependent range of
aggregation and the transition broadness change simultaneously and
consistently with the spacing of stickers. It is demonstrated that the
broad nature of the transitions observed in PAPSs is concerned with
the penetration of solvents from the aggregates, which is in
reasonable agreement with the experimental result observed by
Goldmints et al. in the unimer-micelle transition~\cite{Gold1997}.
At the same time, it is found that the symmetry of a specific
heat peak is affected by the process of penetration of solvents.
When the transition broadness increases, the symmetry of a transition
peak decreases. Furthermore,  from the above behaviors of
penetration of a solvent and heat capacity, it is seen that an increase of spacing of
stickers has different effect on the HS-MFH and MFH-micelle
transitions.

In order to interpret the different effect of spacing of stickers on
the aggregation of stickers in MFH and micelle morphologies, we
evaluate the probability that a sticker of polymer chain forms
intrachain and interchain associations in the system using an
approach similar to the one presented in
reference~\cite{Bras2009,Mats1999,Han2010}. We suppose that there are no
other sticker aggregates in the MFH and micellar system except
sticker-rich site microfluctuations  or the micellar cores. A
sticker in a particular chain can form an intrachain association, as
well as an interchain association. Ignoring the probabilities that more
than two stickers of a definite chain are attached to an aggregate,
the conditional probability that the sticker $s_{1}$ is concerned
with intrachain association, provided that the sticker $s_{1}$ is at
an aggregate of the two above mentioned types whose position locates at
$r_\mathrm{ag}$, can be expressed as:
\begin{equation}
p_\mathrm{loop}(r_\mathrm{ag},s_{1})=\frac{1}{P^{(1)}(r_\mathrm{ag},s_{1})}\sum_{{s_{2}\in \mathrm{st},}%
s_{2}\neq s_{1}}P^{(2)}(r_\mathrm{ag},s_{1};r_\mathrm{ag},s_{2}),
\end{equation}%
where $\sum_{{s_{2}\in \mathrm{st},s_{2}\neq s_{1}}}$ means the summation
over all
the stickers of a polymer chain except the $s_{1}$th one, while $P^{(1)}(r_\mathrm{ag},s_{1})$ and $P^{(2)}(r_\mathrm{ag},s_{1};r_\mathrm{ag},s_{2})$,
whose expressions are given in appendix, are the single-segment and
two-segment probability distribution functions of a chain,
respectively. Then, $1-p_\mathrm{loop}(r_\mathrm{ag},s_{1}) $ is the conditional
probability that the sticker $s_{1}$ is linked with those that
belong to other chains when the sticker $s_{1}$ is at $r_\mathrm{ag}$, and%
\[
\mathbf{P}_{lk}(s_{1})=\sum_{r_\mathrm{ag}}P^{(1)}(r_\mathrm{ag},s_{1})
\left[ 1-p_\mathrm{loop}(r_\mathrm{ag},s_{1})\right]
\]%
is the probability that a sticker $s_{1}$ of a chain is related to
an interchain association, where $\sum_{r_\mathrm{ag}}$ means the summation
over all the aggregates in the system. The summation of
$\mathbf{P}_{lk}(s_{1})$ over all the stickers in a chain, $\langle
n_{_{lk}}\rangle =\sum_{s_{1},s_{1}\in \mathrm{st}}\mathbf{P}_{lk}(s_{1})$,
can be viewed as the average sticker number from a particular
polymer chain linked with other chains by sticker aggregates. The
average fraction of interchain association of a sticker
 is expressed as $\bar{f}_{te}=\langle
n_{_{lk}}\rangle/N_\mathrm{st}$. The average fraction of intrachain
association of a sticker
 is defined as $\bar{f}_\mathrm{tr} =(1/N_\mathrm{st})\sum_{s_{1},s_{1}\in
\mathrm{st}}\sum_{{r_\mathrm{ag}}}p_\mathrm{loop}(r_\mathrm{ag},s_{1})$.

\begin{figure}[ht]
\includegraphics[width=0.48\textwidth]{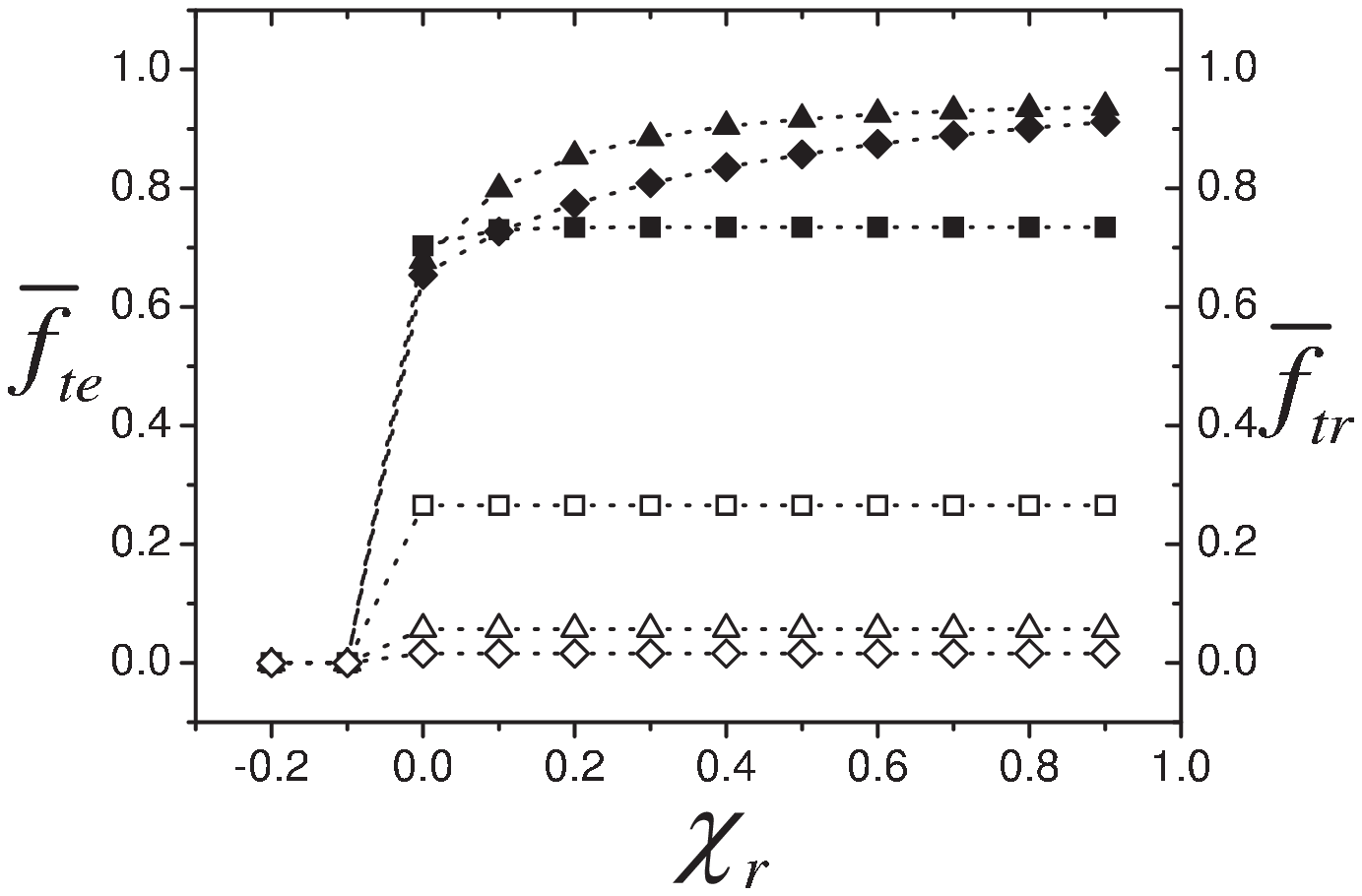}%
\hfill%
\includegraphics[width=0.48\textwidth]{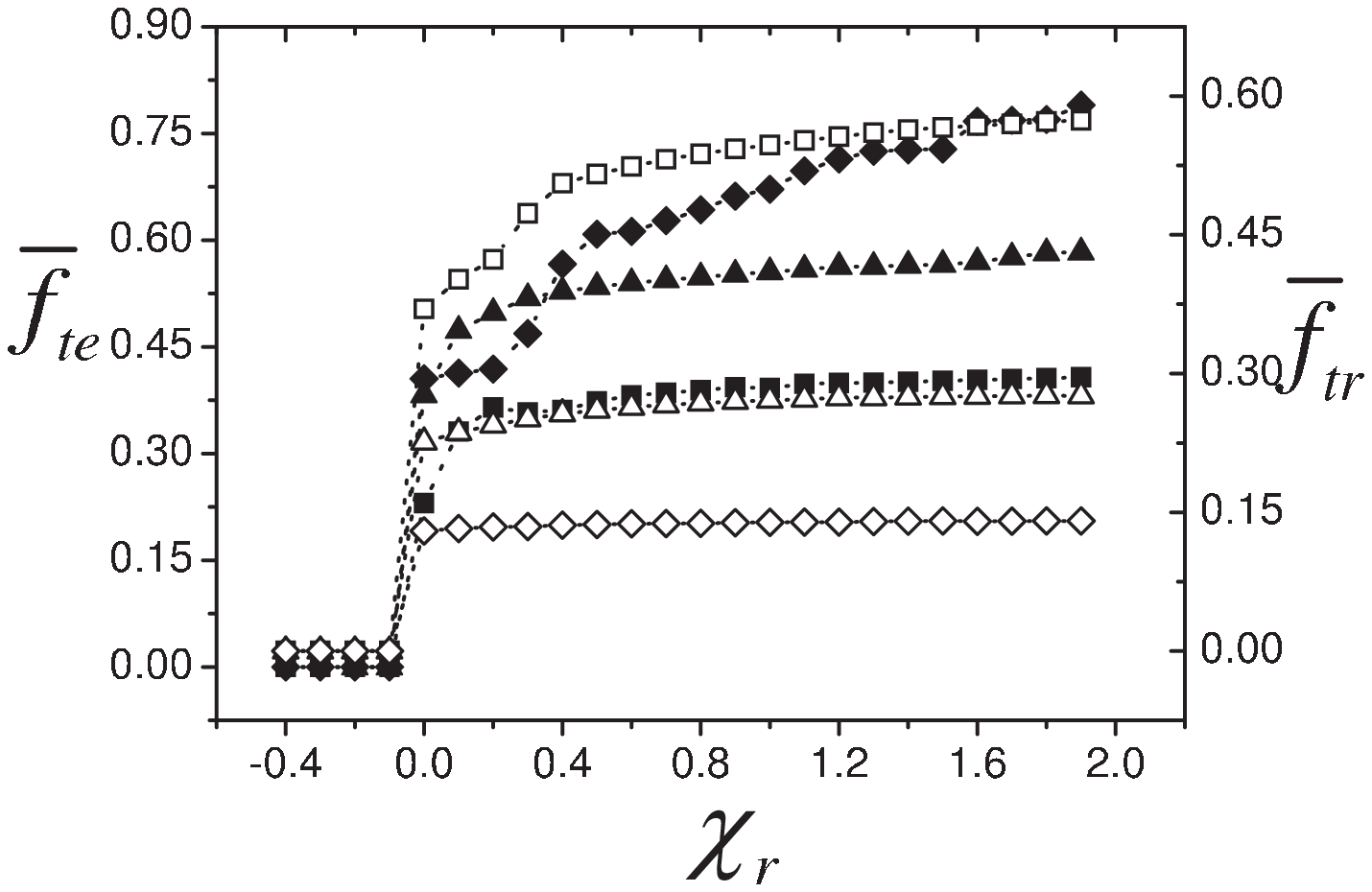}%
\\%
\parbox[t]{0.48\textwidth}{%
\centerline{(a)}%
}%
\hfill%
\parbox[t]{0.48\textwidth}{%
\centerline{(b)}%
}%
\caption{The average fractions of intrachain and interchain
associations of a sticker, denoted by $\bar{f}_\mathrm{tr}$ and
$\bar{f}_\mathrm{te}$, respectively, as functions of the $\chi$ deviation
from MFH or micellar boundary $\chi_\mathrm{r}$ are presented in
figures~\ref{loop}~(a) and (b) in MFH and micelle morphologies in the
systems with different spacing of stickers $l$ at
$\bar{\phi}_\mathrm{P}=0.8$. The red open and solid squares, green open
and solid triangles, blue open and solid diamonds correspond to by
$\bar{f}_\mathrm{tr}$ and $\bar{f}_\mathrm{te}$ of $l=3, 9, 19$, respectively.}
\label{loop}
\end{figure}

Figure~\ref{loop} shows the variations of average fractions of
intrachain association and interchain association of a sticker,
denoted by $\bar{f}_\mathrm{tr}$ and $\bar{f}_\mathrm{te}$, respectively, with the
$\chi$ deviation from the MFH or micellar boundary, $\chi_\mathrm{r}$, in the
system with different spacing of stickers $l$ at
$\bar{\phi}_\mathrm{P}=0.8$. In MFH morphology [figure~\ref{loop}~(a)],
when $\chi_\mathrm{r}$ is increased, $\bar{f}_\mathrm{te}$ at $l=3$ first rises when
$0\leqslant\chi_\mathrm{r}\leqslant0.2$, then maintains a certain value, and the
corresponding $\bar{f}_\mathrm{tr}$ always remains constant when
$\chi_\mathrm{r}>0$, and $\bar{f}_\mathrm{tr}$ is smaller than $\bar{f}_\mathrm{te}$. When
$l$ is increased, the variation of $\bar{f}_\mathrm{te}$ with $\chi_\mathrm{r}$
resembles that of $l=3$. However, a certain value  at which
$\bar{f}_\mathrm{te}$ finally arrives rises, and the corresponding $\chi_\mathrm{r}$
also goes up, with  $\chi_\mathrm{r}$ increasing. $\bar{f}_\mathrm{tr}$ decreases
markedly when $l$ is increased. $\bar{f}_\mathrm{tr}$ is much smaller than the
corresponding $\bar{f}_\mathrm{te}$. It is shown that in MFH morphology
intrachain association is independent of $\chi_\mathrm{r}$, and the range of
interchain association which is concerned with $\chi_\mathrm{r}$ rises when
spacing of stickers is increased. In micelle morphology [figure~\ref{loop}~(b)], the variation of the average fraction of interchain
association of a sticker $\bar{f}_\mathrm{te}$ with the deviation from
micellar boundary $\chi_\mathrm{r}$ and the effect of $l$ on it are similar
to those of MFH morphology. However, there exists an evident
difference in the curve of $\bar{f}_\mathrm{te}(\chi_\mathrm{r})$, which is not
smooth.
The intrachain associations of micelles $\bar{f}_\mathrm{tr}$ are dependent on  $\chi_\mathrm{r}$. The behavior is distinct for small $l$ and different from MFH morphology. It is noted that at $l =3$, when $0<\chi_\mathrm{r}\leqslant0.4$, the extent of rise of $\bar{f}_\mathrm{tr}$ and $\bar{f}_\mathrm{te}$  changes alternately with increasing $\chi_\mathrm{r}$.

\looseness=-1As discussed above, the increase of spacing of stickers $l$ has
different effect on the HS-MFH and MFH-micelle transitions, which
can be explained in terms of intrachain and interchain associations.
The average fraction of interchain association of a sticker in MFH
morphology $\bar{f}_\mathrm{te}^\mathrm{ri}$ is much larger than that of
the corresponding intrachain quantity $\bar{f}_\mathrm{tr}^\mathrm{ri}$, and
$\bar{f}_\mathrm{tr}^\mathrm{ri}$ is absolutely independent of temperature.
Therefore, the temperature-dependent property of MFH morphology is
determined by interchain association. When $l$ is increased,
$\bar{f}_\mathrm{te}^\mathrm{ri}$ range concerned with  temperature also rises.
Therefore, an increase of spacing of stickers is favorable to an
increase in the broadness of HS-MFH transition. Meanwhile, in micelle
morphology, the average fraction of intrachain association of a
sticker $\bar{f}_\mathrm{tr}^\mathrm{ro}$ is dependent on  temperature, especially
in the case of small $l$. Therefore, the property of MFH-micelle
transition is determined by both intrachain and interchain
associations [figure~\ref{loop}~(b)]. At $l=3$, both
$\bar{f}_\mathrm{tr}^\mathrm{ro}$ and $\bar{f}_\mathrm{te}^\mathrm{ro}$ are sensitive to
 temperature in the transition region. When $l$ is increased, the
susceptibility of $\bar{f}_\mathrm{tr}^\mathrm{ro}$ to  temperature decreases
markedly. Although an increase in $l$ is favorable to an increase
in the susceptibility of $\bar{f}_\mathrm{te}^\mathrm{ro}$ to  temperature, it may
be weak compared with the corresponding decrease of
$\bar{f}_\mathrm{tr}^\mathrm{ro}$. Therefore, when spacing of stickers is
increased, the broadness of the corresponding MFH-micelle transition
decreases, which is different from that of HS-MFH transition.

\section{Conclusion and summary\label{sec4}}

The effect of distribution of stickers along the backbone on the
temperature-dependent property of aggregation structure in
physically associating polymer solutions (PAPSs) is studied using
the self-consistent field lattice model. When spacing of stickers is
increased, the temperature susceptibility of penetration of
solvents from aggregates in MFH morphology and  the broadness of
HS-MFH transition increases. However, the corresponding two
quantities of MFH-micelle transition do decrease under the same
condition, which is opposite to that of HS-MFH transition. It is
found that the temperature susceptibility of penetration of
solvents from the two above morphologies and the broadness of the
two transitions change simultaneously and consistently. It is demonstrated that the broadness of the transitions observed in
PAPSs is concerned with the penetration of solvent from aggregates.
Furthermore, the different effect of spacing of stickers on HS-MFH
and MFH-micelle transitions is due to different contributions of
intrachain and interchain associations to MFH and micelle
morphologies. This work can be extended to the study of the effects of
polymer concentration and chain architecture on the transition
properties related to the penetration of a solvent.

\section*{Acknowledgements} This research is financially supported by the National Nature Science Foundations of China (11147132) and the Inner Mongolia municipality (2012MS0112), and the Innovative Foundation of Inner Mongolia University of
Science and Tech\-no\-lo\-gy (2011NCL018).

\section*{Appendix}
Following the scheme of Schentiens and Leermakers~\cite{Leer1988},
$G^{\alpha _{s}}(r,s|1)$ is the end segment distribution function of
the $s$th segment of the chain, which is evaluated from the
following recursive relation:
\begin{equation} G^{\alpha
_{s}}(r,s|1)=G(r,s)\sum_{r_{s-1}'}\sum_{\alpha
_{s-1}}\lambda _{r_{s}-r_{s-1}'}^{\alpha _{s}-\alpha
_{s-1}}G^{\alpha _{s-1}}(r',s-1|1), \label{free}
\end{equation}
where $G(r,s)$ is the free segment
weighting factor and is expressed as
\[
G(r,s)=\left\{
\begin{array}{ll}
\exp[-\omega _\mathrm{st}(r_{_{s}})],\qquad & s\in
\mathrm{st}\,, \\
\exp[-\omega_\mathrm{ns}(r_{_{s}})],\qquad&  s\in \mathrm{ns}\,.
\end{array}\right.
\]
The initial condition is $G^{\alpha _{1}}(r,1|1)=G(r,1)$ for
all the values of $\alpha _{1}$. In the above expression, the values
of $\lambda _{r_{s}-r'_{s-1}}^{\alpha _{s}-\alpha _{s-1}}$ depend
on the chain model used. We assume that
\[
\lambda
_{r_{s}-r'_{s-1}}^{\alpha _{s}-\alpha _{s-1}}=\left\{
                                                   \begin{array}{ll}
                                                     0, & \alpha _{s}=\alpha _{s-1}\,, \\
                                                     {1}/(z-1), & \textrm{otherwise}.
                                                   \end{array}
                                                 \right.
\]
This means that the chain is described as a random walk without the
possibility of direct backfolding. Another end segment distribution
function $G^{\alpha _{s}}(r,s|N)$ is evaluated from the following
recursive relation: \begin{equation} G^{\alpha
_{s}}(r,s|N)=G(r,s)\sum_{r_{s+1}'}\sum_{\alpha
_{s+1}}\lambda _{r_{s+1}'-r_{s}}^{\alpha _{s+1}-\alpha
_{s}}G^{\alpha _{s+1}}(r',s+1|N),
\end{equation}
with the initial condition $G^{\alpha _{N}}(r,N|N)=G(r,N)$ for all
the values of $\alpha _{N}$.

Using the expressions of the end segment distribution functions, the
single-segment probability distribution function $P^{(1)}(r,s)$ and
the
two-segment probability distribution function $%
P^{(2)}(r_{1},s_{1};r_{2},s_{2})$ of the  chain can be defined as
follows:
\begin{equation}
P^{(1)}(r,s)=\frac{1}{zN_{L}Q_\mathrm{P}}\sum_{r_{s}'}\sum_{{\alpha }%
_{_{s}}}\frac{G^{\alpha _{s}}(r',s|1)G^{\alpha
_{s}}(r',s|N)}{G(r',s)}\delta _{r_{s}',{r}}\, ,
\end{equation}%
which is a normalized probability that the monomer $s$ of the
chain is on the
lattice site $r$;%
\begin{eqnarray}
P^{(2)}(r_{1},s_{1};r_{2},s_{2}) &=&\frac{1}{zN_{L}Q_\mathrm{P}}%
\sum_{r_{s_{1}}'}\sum_{{\alpha }_{_{s_{1}}}}\sum_{r_{s_{2}}'}\sum_{{\alpha }_{_{s_{2}}}}G^{\alpha _{s_{1}}}(r',s_{1}|1)\delta _{{r}_{s_{1}}',{r}_{1}} \\
&&{}\times\mathcal{G}(r',s_{1};r',s_{2})G^{\alpha
_{s_{2}}}(r',s_{2}|N)\delta _{{r}_{s_{2}}',{r}_{2}} \nonumber
\end{eqnarray}%
and
\[
\mathcal{G}(r,s_{1};r,s_{2})=\sum_{r_{s_{1}+1}}\sum_{\alpha
_{s_{1}+1}}\ldots\sum_{r_{s_{2}-1}}\sum_{\alpha _{s_{2}-1}}\left\{
\prod_{s=s_{1}+1}^{s_{2}-1}\lambda _{r_{s}-r_{s-1}}^{\alpha
_{s}-\alpha _{s-1}}G(r,s)\right\} \lambda
_{r_{s_{2}}-r_{s_{2}-1}}^{\alpha _{s_{2}}-\alpha _{s_{2}-1}}\ \ \ \
\ (\textrm{for}\ s_{2}>s_{1})
\]%
give the probability that the monomers $s_{1}$ and $s_{2}$ of
the chain are on the lattice sites $r_{1}$ and $r_{2}$,
respectively. It can
be verified that $\sum_{r}P^{(1)}(r,s)=1$, and $%
\sum_{r_{2}}P^{(2)}(r_{1},s_{1};r_{2},s_{2})=P^{(1)}(r_{1},s_{1})$.

\newpage

\ukrainianpart

\title{Вплив розподілу центрів зв'язування вздовж головного ланцюга на температурно залежні структурні  властивості в асоціативних полімерних розчинах}
\author{К.-Г. Ган, К.-Ф. Жанг, Й.-Г. Ма}

\address{
Школа математики, фізики і біології, університет науки і технологій Внутрішньої Монголії, Баоту 014010, Китай }

\makeukrtitle

\begin{abstract}
\tolerance=3000%
Вплив розподілу  центрів зв'язування (stickers) вздовж  головного ланцюга на  структурні  властивості в асоціативних полімерних розчинах
вивчається з використанням ґраткової моделі самоузгодженого поля. Виявлено лише дві неоднорідні морфології, а саме мікрофлуктуаційну гомогенну (МФГ) та міцелярну морфології. Якщо система є охолодженою, тоді зменшується вміст розчинника всередині агрегатів. Коли відстань між  центрами зв'язування вздовж головного ланцюга збільшується,  тоді зростають температурно залежний діапазон агрегації в морфології МФГ і півширина піку питомої теплоємності для переходу гомогенні розчини-МФГ, а симетрія  піку -- зменшується. Проте, з ростом відстані між  центрами зв'язування вище згадані три величини, пов'язані з міцелами,
поводять себе інакше. Показано, що різний характер спостережених переходів може пояснюватися структурними змінами, які супроводжують заміну розчинників в агрегатах на полімери, що узгоджується з результатами експерименту. Знайдено, що вплив відстані між  центрами зв'язування на ці два переходи можна також  трактувати на мові внутрішньоланцюгових і міжланцюгових асоціацій.
\keywords структурні властивості, самоузгоджене поле, асоціативний полімер

\end{abstract}

\end{document}